\documentclass{webofc}
\usepackage[varg]{txfonts}   
\woctitle{?????????}
\begin{document}
\title{Magnetic fields of Herbig Ae/Be stars }
%
%

\author{S. Hubrig\inst{1}\fnsep\thanks{\email{shubrig@aip.de}}
\and
       I. Ilyin\inst{1}
\and
       M. Sch\"oller\inst{2}
\and
         C. R. Cowley\inst{3}
\and
         F. Castelli\inst{4}
\and
        B. Stelzer\inst{5}
\and
       J. -F. Gonzalez\inst{6}
\and
       B. Wolff\inst{2}
}

\institute{Leibniz-Institut f\"ur Astrophysik Potsdam (AIP), An der Sternwarte 16, 
14482 Potsdam, Germany
\and
 European Southern Observatory, Karl-Schwarzschild-Str.\ 2, 85748 Garching bei M\"unchen, 
Germany         
\and
Department of Astronomy, University of Michigan, Ann Arbor, MI 48109-1042, USA
\and
 Istituto Nazionale di Astrofisica, Osservatorio Astronomico di Trieste, via Tiepolo 11, 34143 
Trieste, Italy
\and  
 INAF-Osservatorio Astronomico di Palermo, Piazza del Parlamento 1, 90134 Palermo, Italy  
\and 
Instituto de Ciencias Astronomicas, de la Tierra, y del Espacio (ICATE), San Juan, Argentina
          }

\abstract{%
 We report on the status of our spectropolarimetric studies of Herbig Ae/Be stars carried out 
during the last years. The magnetic field geometries of these stars, investigated with 
spectropolarimetric time series, can 
likely be described by centred dipoles with polar magnetic field strengths of several hundred 
Gauss. A number of Herbig Ae/Be stars with detected magnetic fields have recently been 
observed with X-shooter in the visible and the near-IR, as well as with 
the high-resolution near-IR spectrograph CRIRES. These observations are of great importance 
to understand the relation between the magnetic field topology and 
the physics of the accretion flow and the accretion disk gas emission.
 }
\maketitle

\section{Introduction}
\label{sect:intro}

Magnetic fields are important ingredients of the star formation process (e.g.\ McKee \& Ostriker 2007 \cite{McKeeOstriker2007}).
Models of magnetically driven accretion and outflows successfully reproduce many 
observational properties of low-mass pre-main sequence stars. Indirect observational evidence for 
the presence of magnetic fields in these stars is manifested in strong X-ray, FUV, and UV emission 
(e.g.\ Feigelson \& Montmerle 1999 \cite{FeigelsonMontmerle1999}).
The first detections of magnetic fields in protostars of class I and II sources were obtained 
using NIR spectrographs and revealed kG fields (e.g.\ Johns-Krull et al.\ 2009 \cite{JohnsKrull2009}).  
The first 
magnetic-field maps of T\,Tauri stars show some systems that have complex fields while some 
have much simpler dipolar/octupolar fields (e.g.\ Donati et al.\ 2008 \cite{Donati2008}). 
Accretion models based on these maps demonstrate 
the strong dependence of accretion efficiency on both the strength and the geometry of the star’s 
magnetic field.  

Fields have also been detected in a dozen Herbig Ae/Be stars 
(e.g.\ Hubrig et al.\ 2009 \cite{Hubrig2009}). Similar to T\,Tau stars, Herbig Ae/Be stars show 
clear signatures of 
surrounding disks as evidenced by a strong infrared excess and are actively accreting material.  
 Current theories are not able to present a consistent scenario of how the magnetic fields in 
Herbig Ae/Be stars are generated and how these fields interact 
with the circumstellar environment, consisting of a combination of disk, wind, accretion, and jets. 
On the other hand, understanding the interaction between the central stars, their
magnetic fields, and their protoplanetary disks is crucial for reconstructing the Solar
System's history, and to account for the diversity of exo-planetary systems.

\section{Recent studies of the presence of magnetic fields in Herbig Ae/Be stars}
\label{sect:studies}

Before 2004, the only magnetic field detection of about 50\,G had been reported for the 
optically brightest 
($V=6.5$) Herbig Ae star HD\,104237 (Donati et al.\ 1997 \cite{Donati1997}), but no further publications 
confirming this detection existed until recently. Using high-resolution, high 
signal-to-noise HARPS\-pol observations 
(Hubrig et al.\ 2013 \cite{Hubrig2013}) detected a mean longitudinal magnetic field of the order of 60\,G. 
Spectropolarimetric studies from 2004 to 2008 reported the discovery of magnetic fields in seven other 
Herbig Ae/Be stars
(Wade et al.\ 2005 \cite{Wade2005}, 2007 \cite{Wade2007},
Catala et al.\ 2007 \cite{Catala2007},
Hubrig et al.\ 2004 \cite{Hubrig2004}, 2006 \cite{Hubrig2006}, 2007 \cite{Hubrig2007}).
Later on, a study of 21 Herbig Ae/Be stars with FORS\,1 revealed the presence of magnetic 
fields in six additional stars (Hubrig et al.\ 2009 \cite{Hubrig2009}).
More recent studies involved the outbursting binary Z\,CMa (Szeifert et al.\ 2010 \cite{Szeifert2010}), 
the Herbig Ae star HD\,101412 with resolved magnetically split lines, and HD\,31648\,=\,MWC\,480 
(Hubrig et al.\ 2010 \cite{Hubrig2010}, 2011a \cite{Hubrig2011a}).

\subsection{Magnetic field versus observed properties}
\label{sect:prop}

Spectropolarimetric observations of a sample of 21 Herbig Ae/Be stars observed with FORS\,1
have been used to search for a link between the presence of a magnetic field and other stellar
properties (Hubrig et al.\ 2009 \cite{Hubrig2009}). 
This study did not indicate any correlation of the strength of the longitudinal magnetic 
field with disk orientation, disk geometry, or the presence of a companion. 
No simple dependence on the mass-accretion rate was found, but the range of observed field values 
qualitatively supported the expectations from magnetospheric accretion models with dipole-like 
field geometries. Both the magnetic field strength and the X-ray emission showed hints of 
a decline with age in the range of $\sim2-14$\,Myr probed by the sample, supporting a dynamo 
mechanism that decays with age.  Furthermore, the stars seemed to obey the universal power-law 
relation between magnetic flux and X-ray luminosity established for the Sun and main-sequence 
active dwarf stars.

\subsection{Magnetic field geometry}
\label{sect:geometry}
Series of mean longitudinal magnetic-field measurements were recently obtained at low 
resolution with the multi-mode instrument 
FORS\,2 at the VLT for the Herbig Ae/Be stars HD\,97048, HD\,101412,
HD\,150193, and HD\,176386 (Hubrig et al.\ 2011b \cite{Hubrig2011b}).
Magnetic fields of the order of 120--250\,G were for the first time detected in these stars a 
few years ago during our visitor run 
with FORS\,1 in May 2008 (Hubrig et al.\ 2009 \cite{Hubrig2009}).
In our observations, the Herbig Ae/Be stars exhibit a single-wave variation in the longitudinal 
magnetic field 
during the stellar rotation cycle. This behaviour is usually considered as evidence 
for a dominant dipolar contribution to the magnetic field topology.
Presently, the Herbig Ae star HD\,101412 possesses the strongest 
longitudinal magnetic field ever measured in any Herbig Ae star, 
with a surface magnetic field $\left<B\right>$ up to 3.5\,kG. 
HD\,101412 is also the only Herbig Ae/Be star for which the rotational
Doppler effect was found to be small in comparison to the magnetic splitting 
and  several spectral lines
observed in unpolarised light at high dispersion are resolved into 
magnetically split components (Hubrig et al.\ 2010 \cite{Hubrig2010}, 2011b \cite{Hubrig2011b}).

\begin{figure*}
\centering
\includegraphics[width=0.9\textwidth]{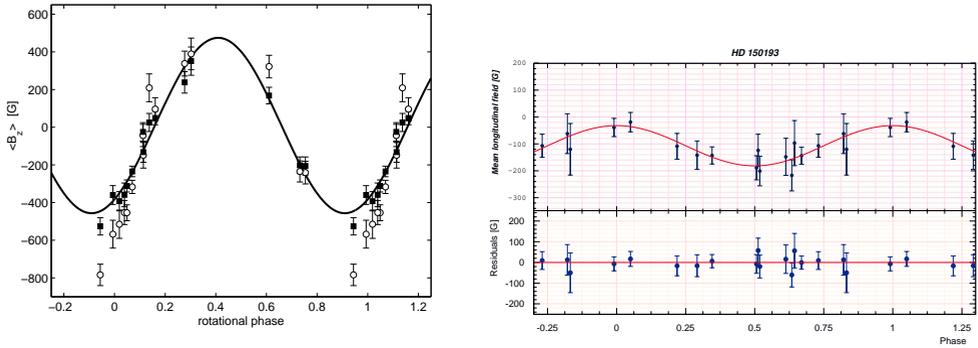}
\caption{
Left panel: Phase diagram of HD\,101412 with the best sinusoidal fit for the longitudinal magnetic field measurements
using all lines (filled squares) and hydrogen lines (open circles). Right panel: Phase diagram of HD\,150193
with the best sinusoidal fit for the longitudinal magnetic field measurements
using all lines. The residuals (Observed -- Calculated) are shown in the lower panel.
}
\label{fig:1}
\end{figure*}

To date, magnetic field geometries have been studied for the two SB2 systems HD\,200775 (with a B3 primary) 
and V380\,Ori (with a B9 primary) (Alecian et al.\ 2008 \cite{Alecian2008}, 2009 \cite{Alecian2009}),
and the presumably single stars HD\,101412, HD\,97048, HD\,150193, and HD\,176386
(Hubrig et al.\ 2010 \cite{Hubrig2010}, 2011b \cite{Hubrig2011b}).
As an example, phase diagrams of the magnetic data for the Herbig Ae/Be stars
HD\,101412 and HD\,150193 folded with the determined magnetic/rotation periods are 
presented in Fig.~\ref{fig:1}.

Our magnetic field model 
for the Herbig Ae star HD\,101412 is described 
by a centered dipole with a polar magnetic field strength $B_{\rm d}$ between
1.5 and 2\,kG and an inclination of the magnetic axis to the rotation axis $\beta$ of 
84$\pm$13$^{\circ}$ (Hubrig et al.\ 2011a \cite{Hubrig2011a}).
The fact that the dipole axis is located close to the stellar equatorial 
plane is very intriguing in view of the generally assumed magnetospheric accretion scenario that 
magnetic fields channel the accretion flows towards the stellar surface along magnetic field lines.
As was shown in the past (Romanova et al.\ 2003 \cite{Romanova2003}),
the topology of the channeled accretion critically depends on the 
tilt angle between the rotation and the magnetic axis. For large inclination angles $\beta$,
many polar field lines would thread the inner region of the disk, while the closed 
lines cross the path of the disk 
matter, causing strong magnetic braking, which could explain the observed unusually 
long rotation period of HD\,101412 of about 42 days.

Since about 70\% of the Herbig Ae/Be stars appear in binary/multiple systems
(Baines et al.\ 2006 \cite{Baines2006}), special
care has to be taken in assigning the measured magnetic field to the particular component in the 
Herbig Ae/Be system.
Alecian et al.\ (2009 \cite{Alecian2009}) reported on the discovery of a dipolar magnetic field in the Herbig Be star
HD\,200775, which is a double-lined spectroscopic binary system. However, it should be noted that the 
magnetic field was discovered not in the component possessing a circumstellar disk and 
dominating the H$\alpha$ emission, so that the evolutionary status of the B3 primary component 
is yet unclear (Benisty et al.\ 2013 \cite{Benisty2013}). 
Similar to the case of HD\,200775, the frequently mentioned discovery of a 
magnetic field in the Herbig SB2 system 
HD\,72106  (Alecian et al.\ 2009 \cite{Alecian2009}) refers to the detection only in 
the primary component, which is a young main-sequence star, but not in the Herbig Ae secondary 
(Folsom et al.\ 2008 \cite{Folsom2008}).
The same uncertainty in the evolutionary status applies to the magnetic field detection
in the system  V380\,Ori reported by Alecian et al.\ (2009 \cite{Alecian2009}).
The authors detected the presence of a dipole magnetic field of 
polar strength $2.12 \pm 0.15$\,kG on the surface of the chemically peculiar primary 
V380\,Ori system.
V380\,Ori has a spectral type around B9 and 
has been observed in great detail over many wavelength ranges
(e.g., Hamann \& Persson 1992 \cite{HamannPersson1992},
Rossi et al.\ 1999 \cite{Rossi1999}, Stelzer et al.\ 2006 \cite{Stelzer2006}).
It has a close infrared companion, with
a separation of 0.15$^{\prime\prime}$ at PA 204$^{\circ}$ (Leinert et al.\ 1997 \cite{Leinert1997}). 
Alecian et al.\ (2009 \cite{Alecian2009}) found that the primary in the 
V380\,Ori system is itself a spectroscopic binary with
a period of 104\,days, with the secondary being a
massive T\,Tauri star. 
Most recently, Reipurth et al.\ (2013 \cite{Reipurth2013}) report that V380\,Ori is
a hier\-ar\-chi\-cal quadruple system with a fourth component at a distance of 8.8$^{\prime\prime}$ and 
position angle 120.4$^{\circ}$.
Since no periodicity was found in the behaviour of the emissions in hydrogen, helium, calcium, 
and oxygen lines (the lines determining the Herbig Ae/Be nature), it is possible that the primary 
chemically peculiar component with the detected 
dipolar magnetic field is already at an advanced age and the Herbig Be 
status of the primary is merely 
based on the appearance of emission in the above mentioned lines belonging to the secondary T\,Tau component.
 
The  presence of a magnetic field on the surface of the Herbig Ae star HD\,190073 is already known for several years.
The first measurement of a longitudinal magnetic field in HD\,190073 was published by 
Hubrig et al.\ (2006 \cite{Hubrig2006}),
indicating the presence of a longitudinal magnetic field $\left<B_{\rm z}\right> =84\pm30$\,G measured 
on FORS\,1 low-resolution spectra at a 2.8$\sigma$ level.  Later on this star was studied by
Catala et al.\ (2007 \cite{Catala2007})
using ESPaDOnS observations, who confirmed the presence of a weak longitudinal magnetic field, 
$\left<B_{\rm z}\right> =74\pm10$\,G, at a higher significance level. 
A few years later a longitudinal magnetic field $\left<B_{\rm z}\right> =104\pm19$\,G  was 
reported by Hubrig et al.\ (2009 \cite{Hubrig2009})
using FORS\,1 measurements. The measurement of 
the longitudinal magnetic field using archival HARPS observations from May 2011, 
$\left<B_{\rm z}\right> =91\pm18$\,G, 
fully confirms the presence of a rather stable weak field (Hubrig et al.\ 2013 \cite{Hubrig2013}). 
Surprisingly, new observations of this star during July 2011 and October 2012 by
Alecian et al.\ (2013 \cite{Alecian2013})
detected variations of the Zeeman signature in the LSD spectra on timescales of days to weeks.
The authors suggest that the
detected variations of Zeeman signatures are the result of the interaction between the fossil 
field and the ignition of a dynamo field generated in the newly-born convective core. As our recent 
measurements (Hubrig et al.\ 2013 \cite{Hubrig2013}) completely
contradict those presented by Alecian et al.\ (2013 \cite{Alecian2013}),
which indicate $\left<B_{\rm z}\right> =-10\pm20$\,G, 
careful spectropolarimetric monitoring over the next years is important to confirm the reported 
variability of the magnetic field.
Furthermore, since HD\,190073 is very likely a binary system
(Baines et al.\ 2006 \cite{Baines2006}),
special care has to be taken in the interpretation of the magnetic field measurements.

\subsection{CRIRES and X-shooter observations of magnetic Herbig Ae/Be stars}
\label{sect:obs}

As mentioned above, all previously studied Herbig Ae/Be stars exhibit a single-wave variation in 
the longitudinal magnetic field during the stellar rotation cycle. These observations are usually 
considered as evidence for a dominant dipolar contribution to the magnetic field topology. 
Magnetospheric accretion theories traditionally consider simple $\sim$kG dipolar magnetic fields
that truncate the disk and force in-falling gas to flow along the field lines.
The assumption of the 
dominance of dipole fields is usually made for simplicity or due to the lack of available 
information
about the true large-scale magnetic field topology of these stars.
Indeed, the recent work of 
Adams \& Gregory (2012 \cite{AdamsGregory2012})
shows that high order field components may even
play a dominant role in the physics of the gas inflow, as the accretion columns 
approach the star.

The rather new diagnostic He\,I $\lambda$~1.083\,$\mu$m 
emission line is considered as probing inflow (accretion) and outflow (winds) in the star-disk 
interaction region of accreting T\,Tauri and Herbig Ae/Be stars.
The uniqueness of this probe derives from 
the metastability of this transition and makes it a good indicator
of wind and funnel flow geometry (Edwards et al.\ 2006 \cite{Edwards2006}).
Further, according to Edwards et al., the He~I 
line appears in emission for higher mass accretion rates and in net absorption for 
lower mass accretion rates.
Modeling of this line allowed Gregory et al.\ (2013, {\sl in preparation}) for the first time to study the 
influence of field topologies on the star-disk interaction. Their models use magnetic 
fields with an observed degree
of complexity, as determined via field extrapolation from stellar magnetic maps.

\begin{figure*}
\centering
\includegraphics[width=0.33\textwidth]{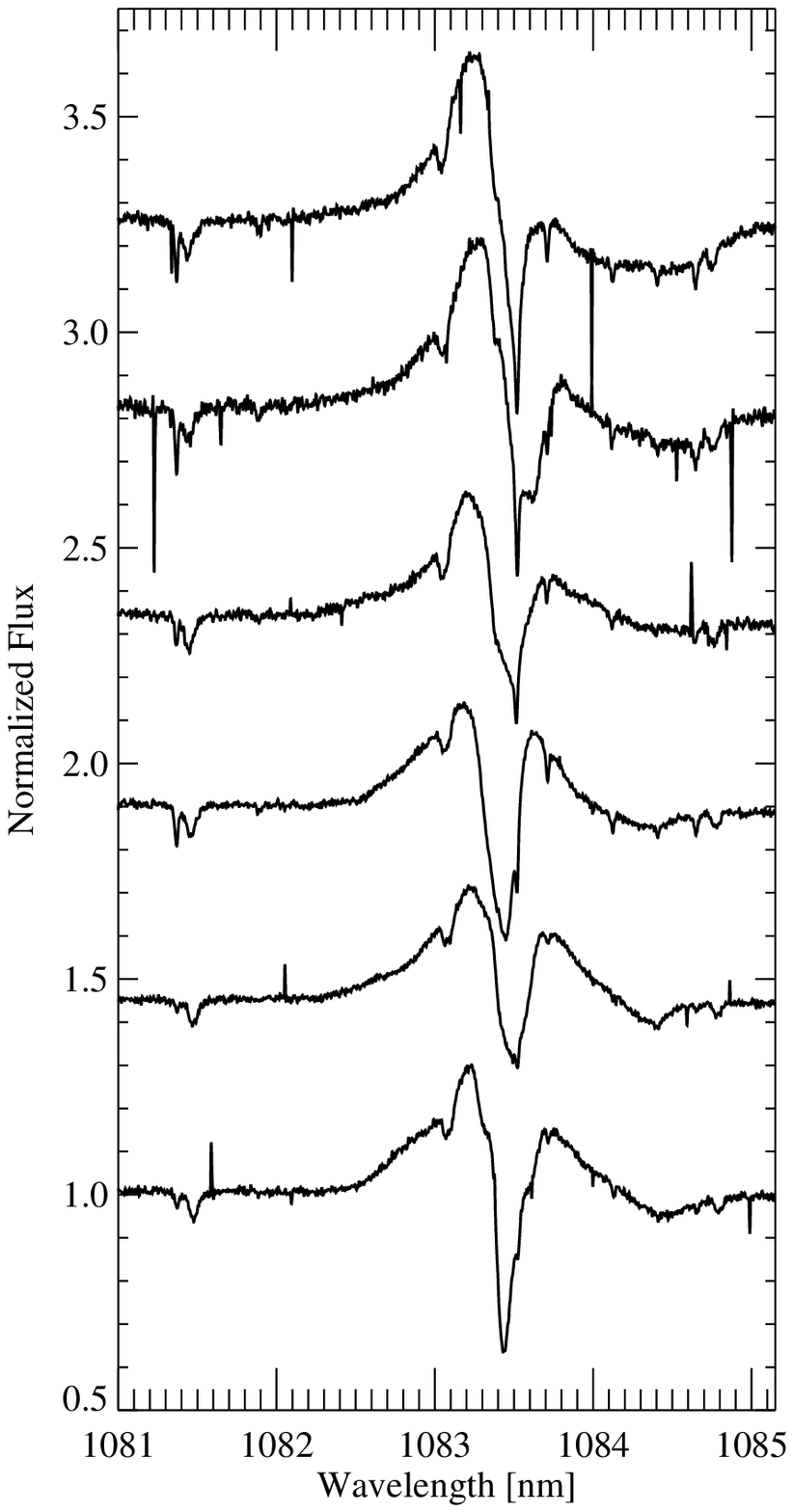}
\includegraphics[width=0.33\textwidth]{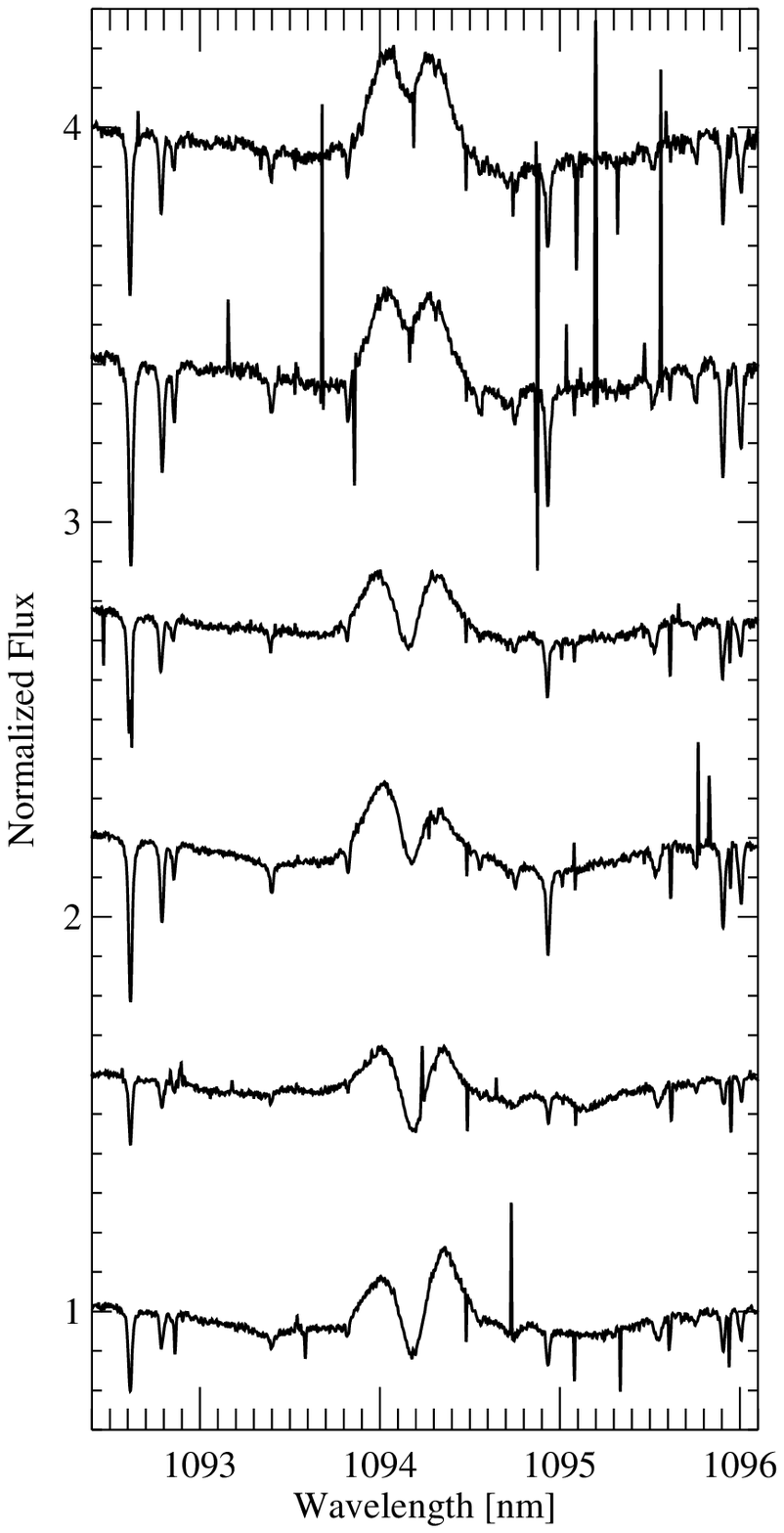}
\caption{
Recent CRIRES observations of HD\,101412. Left panel: The variability 
of the He~I $\lambda$~1.083\,$\mu$m line profile 
over the rotation period.
Obviously, the field of  HD\,101412 appears more complex than just a dipole.
Right panel: Variations of the hydrogen recombination 
line Pa$\gamma$ at 1.094\,$\mu$m at the same rotation phases. The Pa$\gamma$ 
line at 1.094\,$\mu$m is frequently employed for 
calculating the mass accretion rate in the way presented by Gatti et al.\ (2008 \cite{Gatti2008}).
}
\label{fig:2}
\end{figure*}

\begin{figure}
\centering
\includegraphics[width=0.65\textwidth]{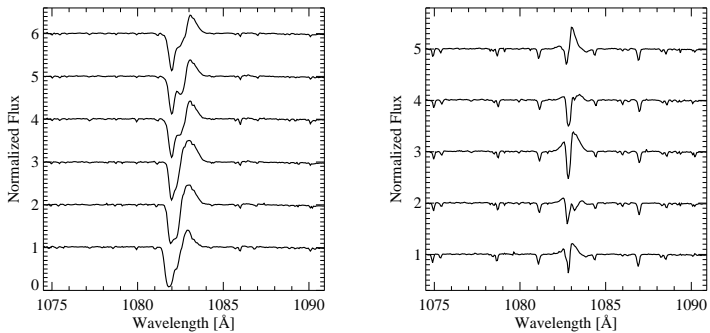}
\caption{
Recent X-shooter observations of the He~I $\lambda$~1.083\,$\mu$m line profile in the 
magnetic Herbig Ae stars HD\,190073 (left panel) and PDS\,2 (right panel) at different epochs. 
The spectra are shifted vertically for clarity.
}
\label{fig:3}
\end{figure}

In Fig.~\ref{fig:2} we present our recent high-resolution CRIRES observations of the spectral regions containing
the He~I $\lambda$~1.083\,$\mu$m line and the hydrogen recombination 
line Pa$\gamma$ at 1.094\,$\mu$m over the rotation period of HD\,101412 (Hubrig et al.\ 2012 \cite{Hubrig2012}). The rather 
strong variation of the line profile
of the He~I line indicates that the magnetic field of this star is likely more 
complex than a dipole field.
A variable behaviour of the He~I $\lambda$~1.083\,$\mu$m line was also discovered in 
our recent 
X-shooter spectra of the magnetic Herbig Ae stars HD\,190073 and PDS\,2 (see Fig.~\ref{fig:3}).

\section{Discussion}
\label{sect:disc}

To understand the magnetospheres of Herbig Ae/Be stars
and their interaction with the circumstellar environment presenting
a combination of disk, wind, accretion, and jets, the
knowledge of the magnetic field strength and topology is indispensable.
Progress in understanding
the disk-magnetosphere interaction can, however,
only come from studying a sufficient number of targets in
detail to look for various patterns encompassing this type of pre-main sequence stars.

\end{document}